\documentclass[aps,twocolumn,groupedaddress,showpacs]{revtex4}
\usepackage{bm}
\begin{document}
%-------------------------------------------------------------------------
% Definitions needed for the heading
%-------------------------------------------------------------------------
\def\Barcelo{Barcel\'o}
\def\Mario{M\'ario}
\def\eg{{\emph{e.g.}}}
%-------------------------------------------------------------------------
\title[Bi-refringence versus bi-metricity.]{Bi-refringence versus bi-metricity.}
%-------------------------------------------------------------------------
\author{Matt Visser}
\email{visser@kiwi.wustl.edu}
\homepage{http://www.physics.wustl.edu/~visser}
\thanks{Supported by the US DOE}
\altaffiliation{Permanent address after 1 July 2002:
School of Mathematics and Computer Science, 
Victoria University, PO Box 600, New Zealand;
{\sf matt.visser@mcs.vuw.ac.nz}}
\affiliation{Physics Department, Washington University, Saint
 Louis MO 63130--4899, USA}
%-------------------------------------------------------------------------
\author{Carlos \Barcelo}
\email{carlos.barcelo@port.ac.uk}
%\homepage{http://www.physics.wustl.edu/~carlos}
\thanks{Supported by the EC under contract HPMF--CT--2001--01203}
%\altaffiliation{}
\affiliation{Institute of Cosmology and Gravitation,
Portsmouth University, Portsmouth PO1 EGR, United Kingdom}
%-------------------------------------------------------------------------
\author{Stefano Liberati}
\email{liberati@physics.umd.edu}
\homepage{http://www2.physics.umd.edu/~liberati}
\thanks{Supported by the US NSF}
%\altaffiliation{}
\affiliation{Physics Department, University of Maryland,
College Park, MD 20742--4111, USA}
%-------------------------------------------------------------------------
\author{\ }
%-------------------------------------------------------------------------
\date{5 February 2002; 
%file {\sf novello.tex}; hep-th/0204017; 
\LaTeX-ed \today}
%-------------------------------------------------------------------------
%-------------------------------------------------------------------------
\begin{abstract}
%-------------------------------------------------------------------------

\bigskip
\centerline{\bf 
Contribution to the Festschrift in honour of Professor {\Mario}
Novello.}

\bigskip

In this article we carefully distinguish the notion of bi-refringence
(a polarization-dependent doubling in photon propagation speeds) from
that of bi-metricity (where the two photon polarizations ``see'' two
distinct metrics). We emphasise that these notions are logically
distinct, though there are special symmetries in ordinary
(3+1)-dimensional nonlinear electrodynamics which imply the stronger
condition of bi-metricity. 

To illustrate this phenomenon we investigate a generalized version of
(3+1)-dimensional nonlinear electrodynamics, which permits the
inclusion of arbitrary inhomogeneities and background fields. [For
example dielectrics (\emph{a l\'a} Gordon), conductors (\emph{a l\'a}
Casimir), and gravitational fields (\emph{a l\'a} Landau--Lifshitz).]
It is easy to demonstrate that the generalized theory is
bi-refringent: In (3+1) dimensions the Fresnel equation, the
relationship between frequency and wavenumber, is always
{\emph{quartic}}. It is somewhat harder to show that in some cases
(\eg, ordinary nonlinear electrodynamics) the quartic factorizes into
two {\emph{quadratics}} thus providing a bi-metric theory. Sometimes
the quartic is a perfect square, implying a single unique effective
metric. We investigate the generality of this factorization process.

%-------------------------------------------------------------------------
\end{abstract}
%---------------------------------------------------------------------
%\pacs{hep-th/0204017.}
%-------------------------------------------------------------------------
\maketitle
%-------------------------------------------------------------------------
% Some local definitions
%---------------------------------------------------------------------
\def\Box{\kern0.5pt{\lower0.1pt\vbox{\hrule height.5pt width 6.8pt
    \hbox{\vrule width.5pt height6pt \kern6pt \vrule width.3pt}
    \hrule height.3pt width 6.8pt} }\kern1.5pt}
%---------------------------------------------------------------------
\def\Schrodinger{Schr\"odinger}
%----------------------------------------------------------------------
\def\half{{1\over2}}
%---------------------------------------------------------------------
\def\im{{\mathrm i}}
\def\a{{\mathbf a}}
\def\d{{\mathrm d}}
\def\m{{\mathrm m}}
%---------------------------------------------------------------------
\def\g{{\mathbf g}}
\def\gi{{\cal G}}
\def\x{{\mathbf x}}
\def\y{{\mathbf y}}
\def\z{{\mathbf z}}
\def\k{{\mathbf k}}
\def\v{{\mathbf v}}
\def\E{{\mathbf E}}
\def\B{{\mathbf B}}
%---------------------------------------------------------------------
\def\*{{}^{\star}\/}
%---------------------------------------------------------------------
\def\*{{}^{\star}\/}
%---------------------------------------------------------------------
\def\cs{c_{\mathrm s}}
\def\ce{c_{\mathrm{eff}}}
%---------------------------------------------------------------------
\def\L{{\cal L}}
\def\I{{\cal I}}
\def\H{{\cal H}}
%---------------------------------------------------------------------
\def\ie{{\emph{i.e.}}}
\def\eg{{\emph{e.g.}}}
\def\etc{{\emph{etc.}}}
\def\etal{{\emph{et al.}}}
%---------------------------------------------------------------------
\def\Schwinger{{\mathrm{Schwinger}}}
\def\Casimir{{\mathrm{Casimir}}}
\def\Maxwell{{\mathrm{Maxwell}}}
\def\background{{\mathrm{background}}}
\def\electron{{\mathrm{e}}}
\def\phase{{\mathrm{phase}}}
\def\group{{\mathrm{group}}}
\def\signal{{\mathrm{signal}}}
\def\effective{{\mathrm{effective}}}
\def\photon{{\mathrm{photon}}}
\def\NLE{{\mathrm{NLE}}}
\def\aka{{\emph{aka}}}
\def\tr{\,\mathrm{tr}\,}
%------------------------------------------------------------------------
\def\be{\begin{equation}}
\def\ee{\end{equation}}
%------------------------------------------------------------------------
\def\bee{\begin{eqnarray}}
\def\eee{\end{eqnarray}}
%------------------------------------------------------------------------
\def\HRULE{{\bigskip\hrule\bigskip}}
%------------------------------------------------------------------------
\section{Introduction}
%-------------------------------------------------
\label{S:introduction}
%-------------------------------------------------

It is commonly assumed that electromagnetic phenomena (in sharp
contrast with gravitational phenomena) can be described by a linear
field theory even in strong fields. But there are many reasons to
expect that this is not ultimately what happens. Born and
Infeld~\cite{Born-Infeld} proposed that some specific non-linearities
could appear at strong fields preventing the existence of arbitrary
large values for the electric field surrounding a charged point
particle. For their part, Euler and Heisenberg~\cite{Euler-Heisenberg}
(see also Schwinger~\cite{Schwinger}) argued that at high values of
the electric field, the quantum creation and annihilation of
electron-positron pairs would give rise to an effective non-linear
electrodynamic theory.  This effective non-linear electrodynamic
theory can be further modified by externally driven alterations of the
quantum vacuum owing to background electromagnetic~\cite{EMbackground}
and gravitational fields~\cite{Gbackground}, or to geometrical
boundary conditions (Casimir plates, \etc)~\cite{GBCbackground}.  For
some additional important examples see the references
in~\cite{Addbackground}.  Remarkably enough, in most cases these
non-linear electrodynamic theories can be described by the use of
appropriate (non-linear) media, with their associated refractive
indices.

Over the last few years Professor {\Mario} Novello and co-authors have
devoted considerable effort to investigating some of the consequences
of having a non-linear electrodynamic theory. On the one hand, they
have been interested in the possible role of non-linear
electrodynamics in circumventing the singularity problem of general
relativity.  In particular, they have shown that in Euler--Heisenberg
electrodynamics it is possible to have a cosmology without an initial
singularity~\cite{NovelloGS}.  On the other hand, Professor Novello
and co-workers have been investigating what effects one should expect
to see in the propagation of photons; now viewed as linear
perturbations around a background electromagnetic
configuration~\cite{NovelloPP}. They were aware that in this case the
causal propagation of photons is not controlled by the ``physical''
spacetime metric (or others conformally related to it), but by an
``effective'' metric depending on the background electromagnetic
fields~\cite{Plebansky}. Professor Novello and his collaborators
realized that this opened up the possibility of building ``geometric
structures'' in a manner analogous to, but different from, the realm
of usual general relativity. For arbitrary non-linear theories they
have shown that black holes~\cite{NovelloBH},
wormholes~\cite{NovelloWH} and even geometries with closed ``timelike''
curves~\cite{NovelloCTC} can be constructed.  These ``effective''
geometries will only be felt by the photons, while other matter fields
will feel the usual gravitational spacetime metric.

A very important point addressed by Professor Novello and co-workers
is the quite generic appearance of bi-refringence in non-linear
electrodynamics~\cite{NovelloPP}. (See also the work of
Plebanski~\cite{Plebansky}, Dittrich and Gies~\cite{Dittrich-Gies},
Schrodinger~\cite{Schrodinger}, and Boillat~\cite{Boillat}.)  The two
polarization states of the photon propagate differently. The present
authors have similarly been confronted with the question of
bi-refringence (or more generally multi-refringence) in general
systems of second order partial differential equations (PDEs); this
investigation being motivated by an abstract approach to analog
relativity~\cite{normal-modes2}. (The last few years have seen a
proliferation of analog models of/for general relativity in the
literature. See~\cite{Barcelo} for an extensive reference list.) In a
series of papers~\cite{normal-modes,normal-modes2} we have shown that
the crucial issue in building an analog model of general relativity is
the linearization of non-linear field theories around some background
solution. In the present article we will apply the general analysis
and language of~\cite{normal-modes2} to a generalization of non-linear
electrodynamics. We will show that the existence of bi-refringence is
quite easily established, but that the step to bi-metricity (the
existence of two different effective metrics controlling the
propagation of each photon polarization) requires special conditions
that are satisfied by electrodynamics in (3+1) dimensions.

We would also like to point out that nonlinear electrodynamics (in
particular Born--Infeld theory) has in recent years seen a marked
resurgence of interest with the advent of the notion of D-brane (see
for example Polchinski~\cite{Polchinski}). Many physicists feel that
D-branes will be a crucial ingredient in any final formulation of
M/String theory. It happens that the motion of a D-brane in the bulk
spacetime is controlled by a Born--Infeld type action~\cite{Gibbons1}.
This implies that while closed strings propagate following the bulk
spacetime metric, open strings (whose end points are attached to
D-branes) follow an effective metric derived from the Born--Infeld
Lagrangian~\cite{Gibbons2}.

Finally, we point out that much of the formalism developed below owes a
great debt to related work (by SL, Sebastiano Sonego, and MV) on
photon propagation at oblique angles in the Casimir
vacuum~\cite{Oblique}.

%-------------------------------------------------
\section{Generalized Nonlinear Electrodynamics}
%------------------------------------------------
\label{S:general}
%-------------------------------------------------

%
Consider a general class of Lagrangians of the form
\begin{equation}
\label{E:effective-lagrangian} \L_\effective =
\L\left(F_{\mu\nu}(x),B(x)\right).
\end{equation}
Here $F_{\mu\nu}$ denotes the electromagnetic field strength; and it
is assumed that derivatives of this field strength do not occur in the
Lagrangian. In terms of the vector potential
\begin{equation}
F_{\mu\nu}(x) = \partial_\mu A_\nu - \partial _\nu A_\mu.
\end{equation}
In addition $B(x)$ denotes a generic class of {\emph{external}}
non-dynamical background fields.  These could represent, for example,
a refractive index, the 4-velocity of a dielectric, the location and
4-velocity of Casimir plates or other conductors, assorted
inhomogeneities and/or boundary conditions, an external gravitational
field, \etc\ If these background fields are all set to their trivial
position-independent values then the system reduces to ordinary
nonlinear electrodynamics in which the Lagrangian depends only on the
two independent Lorentz invariants that can be constructed from the
field strength tensor (for example, Born--Infeld or Euler--Heisenberg
electrodynamics).

The complete equations of motion for nonlinear electrodynamics
consist of the Bianchi identity,
\begin{equation}
\label{E:bianchi} F_{[\mu\nu,\lambda]}=0,
\end{equation}
plus the dynamical equation
\begin{equation}
\label{E:eom0}
\partial_{\nu} \left( {\partial\L\over\partial F_{\mu\nu} }\right) =0.
\end{equation}
We now adopt a linearization procedure: Split the electromagnetic
field into an internal (possibly dynamical) background field plus a
propagating photon
\begin{equation}
\label{E:linearize} 
F_{\mu\nu}= F_{\mu\nu}^\background + f_{\mu\nu}^\photon.
\end{equation}
Then, assuming the background satisfies the equations of motion and
retaining only linear terms in the propagating photon, we have
\begin{equation}
\label{E:bianchi2} (f^\photon)_{[\mu\nu,\lambda]}=0,
\end{equation}
and
\begin{equation}
\label{E:eom1}
\partial_{\nu}
\left( \left. {\partial^2\L\over\partial F_{\mu\nu} \; \partial
F_{\alpha\beta} } \right|_\background \; f^\photon_{\alpha\beta}
\right) =0.
\end{equation}
On defining
\begin{equation}
\label{E:define-omega} 
\Omega^{\mu\nu\alpha\beta}= \left.
{\partial^2\L\over\partial F_{\mu\nu} \; \partial F_{\alpha\beta}}
\right|_\background,
\end{equation}
equation (\ref{E:eom1}) can be rewritten in the somewhat more
compact form
\begin{equation}
\label{E:eom2}  
\label{E:system}
\partial_\alpha \left(\Omega^{\mu\alpha\nu\beta}\;
f^\photon_{\nu\beta} \right)=0.
\end{equation}
Note that the tensor $\Omega^{\mu\nu\alpha\beta}$ is symmetric with
respect to exchange of the pairs of indices $\mu\nu$ and
$\alpha\beta$, and antisymmetric with respect to exchange of indices
within each pair. That is: $\Omega^{\mu\nu\alpha\beta}$ has most of
the key symmetries of the Riemann tensor. If one wishes to work
directly at the level of the linearized Lagrangian one has:
\be
{\cal L}_{\rm{linearized}} = 
{1\over2} \Omega^{\mu\nu\alpha\beta} \; 
f^\photon_{\mu\nu} \;f^\photon_{\alpha\beta}.
\ee
The key observation is that this linearized Lagrangian generically
leads to birefringence.

We now apply a restricted form of the eikonal approximation by
introducing a slowly varying amplitude $f^{\mu\nu}$ and a rapidly
varying phase $\phi$:
\begin{equation}
\label{E:eikonal} 
f^\photon_{\mu\nu} = f_{\mu\nu}\; {\rm e}^{\im\phi}.
\end{equation}
The 4-wavevector is then defined as $k_{\mu}=\partial_{\mu}\phi$.
This approximation is similar to, but not quite identical with, the
usual eikonal approximation.  This is because one assumes that $\phi$
varies on scales much smaller than those of the background, while, on
the other hand, use of the Lagrangian (\ref{E:effective-lagrangian})
also implies that the components of $k$ are much smaller than the
values fixed by the electron mass. (This is the so-called soft-photon
regime).  Under these hypotheses,
\begin{equation}
\label{E:eom3} 
\Omega^{\mu\alpha\nu\beta}\;k_{\alpha} \;f_{\nu\beta}=0.
\end{equation}
But in general the internal dynamical background field is itself subject to
quantum fluctuations, and to take this into account the coefficients
of this equation are to be identified with the expectation value of
the corresponding quantum operators in the background state
$|\psi\rangle$:
\begin{equation}
\label{E:eom-master} 
\langle\psi | \Omega^{\mu\alpha\nu\beta}|\psi\rangle \;
k_{\alpha} \;f_{\nu\beta}=0.
\end{equation}
In taking this expectation value we are using the fact that the
fluctuations in the internal dynamical background fields are
influenced by the external non-dynamical background fields $B(x)$.
(For example, in the case of the Casimir geometry, by the distance
between the plates.) In the spirit of the restricted eikonal
approximation there is a separation of scales between the internal
dynamical background fluctuations and the propagating photon.

The Bianchi identity (\ref{E:bianchi2}) constrains $f^{\mu\nu}$ to
be of the form
\begin{equation}
\label{E:fgp}
f_{\mu\nu} = k_{\mu} \;a_{\nu}-k_{\nu}\; a_{\mu},
\end{equation}
where we have introduced the linearized gauge potential $a$ for the
propagating field. Inserting (\ref{E:fgp}) into (\ref{E:eom-master})
we find
\begin{equation}
\label{E:eom-master2}
\langle\psi |
\Omega^{\mu\alpha\nu\beta}|\psi\rangle
 \;k_{\alpha} \; k_\beta\;a_\nu=0.
\end{equation}
Note that this last equation implies that any completely antisymmetric
part of $\langle \Omega^{\mu\nu\alpha\beta} \rangle$ can be discarded
without affecting the equations of motion.

Of course, this entire discussion could alternatively be rephrased in
terms of Hadamard's theory of the propagation of weak
discontinuities~\cite{Hadamard}, the formalism preferred by Professor
Novello~\cite{NovelloPP}. An identical equation (relating the
polarization and the wavevector) is encountered.

We emphasise that the discussion in this section has (so far) been
completely independent of the dimensionality of spacetime. If we now
ask how many independent components arise in
\begin{equation}
\langle \psi | \Omega^{\mu\nu\alpha\beta} | \psi\rangle
=
\langle \psi | \Omega^{\,(\,[\mu\nu]\,[\alpha\beta]\,)} \ \psi \rangle
\end{equation}
we find [in (d+1) dimensions]
\begin{equation}
{d(d+1)(d^2+d+2)\over8}.
\end{equation}
In particular in (3+1) dimensions this quantity has 21 independent
components. One of these components can be taken to be the coefficient
of the Levi--Civita tensor, and so does not affect the equations of
motion (\ref{E:eom-master2}). Another component can be interpreted as
the overall scale of $\Omega$, which again does not affect the
equations of motion.  So the number of useful independent components
in $\Omega$ is 19. In contrast, two light cones only specify
$2\times9=18$ components. (Two metrics would specify $2\times 10=20$
components, but in this paper we are only looking at the null cones.)
It is ultimately this close relationship which makes (3+1) dimensions
so special (and tricky).

%-------------------------------------------------
\section{Fresnel equation}
%-------------------------------------------------
\label{S:fresnel}
%-------------------------------------------------

Equation (\ref{E:eom-master2}) represents a condition for $a$ as a
function of $k$ --- it constrains $a$ to be an eigenvector, with
zero eigenvalue, of the $k$-dependent matrix
\begin{equation}
\label{E:matrix}
A^{\mu\nu}(k)= 
\langle\psi |\Omega^{\mu\alpha\nu\beta}|\psi\rangle \; k_\alpha \; k_\beta.
\end{equation}
Any non-zero solution corresponds to a physically possible field
polarization, that can be identified by a unit polarization vector
$\epsilon$ (provided $a$ is not a null vector --- a possibility
that can always be avoided by a suitable gauge choice).

A necessary and sufficient condition for the eigenvalue problem
$A^{\mu\nu}\,a_\nu=0$ to have non-zero solutions is
$\mbox{det}\left(A^{\mu\nu} \right)=0$; however, this gives us no
information at all. Indeed, any $a$ parallel to $k$ is always a
non-zero solution, so the condition $\det\left(A^{\mu\nu}\right)=0$ is
actually an identity. On the other hand, $a\parallel k$ is merely an
unphysical gauge mode that corresponds to $f_{\mu\nu}=0$ by
(\ref{E:fgp}), so we need to find other, physically meaningful,
solutions of the eigenvalue problem. 

To this end, we exploit gauge invariance under $a \to a +\lambda\, k$
and fix a gauge, thus removing the spurious modes.  For the current
subsection, it is particularly convenient to adopt the temporal gauge
$a_0=0$. Then we can define a polarization vector $\epsilon_\mu\equiv
a_\mu/\left(a_\nu a^\nu\right)^{1/2}$, and the eigenvalue problem
$A^{\mu\nu}\,\epsilon_\nu=0$ splits into the equation
\begin{equation}
\label{E:00} 
A^{0i}\,\epsilon_i=0,
\end{equation}
plus the reduced eigenvalue problem
\begin{equation}
\label{E:reduced} 
A^{ij}\,\epsilon_j=0\,.
\end{equation}
The latter admits a nontrivial solution only if
\begin{equation}
\label{E:detA=0} 
\det\left(A^{ij}\right)=0.
\end{equation}
The condition (\ref{E:detA=0}) plays the same role as the Fresnel
equation in crystal optics~\cite{Landau} --- it is a scalar
equation for $k$ and thus gives the dispersion relation for light
propagating in our ``medium''.

%----------------------------------------------------------------
The spatial components of the matrix (\ref{E:matrix}) are
\begin{eqnarray}
A^{ij} &=& 
\omega^2 \; \langle\psi| \Omega^{i0j0}|\psi\rangle + 
\omega k_m \; \langle\psi| \Omega^{i0jm} + \Omega^{imj0}|\psi\rangle 
\nonumber\\
&&
+  k_m k_n \;  \langle\psi|\Omega^{imjn}|\psi\rangle ,
\end{eqnarray}
where $\omega=k_0$.  Let us define the unit vector
$\hat{k}=\vec{k}/|\vec{k}|$. Then the components of $A^{ij}$ in a
basis with one axis directed along $\hat{k}$ are
\begin{eqnarray}
A^{ij} \; \hat{k}_j &=&
 \omega^2 \; \langle\psi| \Omega^{i0j0} |\psi\rangle \; \hat{k}_j +
\omega\; k_m \; \langle\psi|\Omega^{imj0}|\psi\rangle \; \hat{k}_j 
\nonumber 
\\
&=& \omega \left(\omega \;\langle\psi| \Omega^{i0j0}|\psi\rangle \; \hat{k}_j +
\hat{k}_m \; \langle\psi|\Omega^{imj0}|\psi\rangle \; \hat{k}_j  \right)
\nonumber 
\\
&\equiv& \omega \; V^{i}.
\end{eqnarray}
In particular
\begin{equation}
A^{ij}\;  \hat{k}_i \hat{k}_j = \omega^2 \; 
\langle\psi|\Omega^{i0j0}|\psi\rangle \;
\hat{k}_i \hat{k}_j \equiv \omega^2 \; S.
\end{equation}
If we now specialize for definiteness to (3+1) dimensions, then the
matrix $A^{ij}$ has the following structure
\begin{equation}
\label{eq:matr}
  \left(
  \begin{array}{cc}
    \omega^2 \; S & \omega \; V^{J}\\
    \omega \; V^{I} & T^{IJ}
  \end{array}
         \right),
\end{equation}
where $I$ and $J$ label the two directions orthogonal to $\hat{k}$ in
the sense of vector-space duality. (The $V^I$ are linear in the
4-wavenumber $k$, while the $T^{IJ}$ are quadratic in the 4-wavenumber
$k$.) Evaluating the determinant by expanding in the first row or
column, it is easy to see that every term will contain at least two
factors of $\omega$, which establishes
\begin{equation}
\label{eq:poli}
  \mbox{det}\left( A^{ij}\right)=\omega^2 \; {\cal P}_{4}(k),
\end{equation}
where ${\cal P}_{4}(k)$ is a homogeneous fourth-order polynomial in
the 4-wavenumber $k_\mu=(\omega, |\vec{k}|\; \hat k_i)$. While the
determinant is itself a \emph{sextic}, the physically
interesting part is given by the \emph{quartic} ${\cal P}_4$. In fact,
by the rules for partitioning determinants
\begin{equation}
\mbox{det}\left( A^{ij}\right)=
\omega^2 \; S\; \det\left[T^{IJ}- {V^I\;V^J\over S}\right],
\end{equation}
so that ${\cal P}_4$ is effectively a $2\times2$ determinant
\begin{equation}
{\cal P}_{4}(k)= S \; \det\left[T^{IJ}- {V^I\;V^J\over S}\right],
\end{equation}
where the elements of the $2\times2$ matrix are themselves quadratic
in the 4-wavenumber.  This allows us to re-phrase the current analysis
in the language of our more general ``normal modes''
analysis~\cite{normal-modes2} by introducing a matrix $f^{\mu\nu;IJ}$
(extremely similar to but not identical with the related quantity
introduced in that article) and writing
\begin{equation}
{\cal P}_{4}(k)= \det\left[f^{\mu\nu;IJ} \; k_\mu \; k_\nu \right].
\end{equation}
Here the two polarization states take on the role of (dual) ``field
indices'' as discussed in reference~\cite{normal-modes2}. The
determinant is to be taken on the $IJ$ indices.

The upshot of this analysis is that in the most general case there
appear to be four dispersion relations, corresponding the four roots
of the quartic equation
\begin{equation}
{\cal P}_{4}(k)=0.
\label{eq:quart}
\end{equation}
If we write $k=(\omega,\vec k)= (\omega,|\vec{k}|\;\hat k)$ then two
of these roots correspond to propagation in the $+\hat k$ direction,
while the other two correspond to propagation in the $-\hat k$
direction.

Different polarization states are represented by linearly independent
solutions of the eigenvalue problem (\ref{E:reduced}), under the
condition (\ref{E:detA=0}). Thus, the space of polarizations is
exactly two-dimensional.  Since equation (\ref{E:detA=0}) gives rise
to two dispersion relations, the polarization states actually satisfy
two (in general, different) eigenvalue equations,
\begin{equation}
\label{E:nonsing}
\overline{A}_{(r)}^{\;\mu\nu}\;\epsilon^{(r)}_{\;\nu}=0\,,
\end{equation}
where $r=1,2$ labels the dispersion relations and
$\overline{A}^{(r)}_{\mu\nu}$ is obtained from $A_{\mu\nu}$ by
imposing the corresponding condition on $k$ as derived from
equation~(\ref{eq:quart}).  Indeed, suppose you pick a specific
3-direction $\hat k$ and have by some means determined two independent
polarization states $\epsilon^{(r)}_{\;\nu}$, which are
\emph{implicitly} functions of $\hat k$ and the corresponding
solutions $\bar{k}$ of ${\cal P}_{4}(k) =0$, then one can construct a
pair of two-index matrices
\begin{equation}
\label{E:fake}
\overline{\gi}_{(r)}^{\;\mu\nu}=  
\langle\psi |
\Omega^{\mu\alpha\nu\beta}|\psi\rangle
 \;\epsilon^{(r)}_{\;\alpha}  \;\epsilon^{(r)}_{\;\beta}.
\end{equation}
Although these quantities are matrices with the correct index
structure to be interpreted as ``effective metrics'' they are in the
general case implicitly functions of $\bar\omega$,
$\overline{|\vec{k}|}$, and the direction $\hat k$, and so these
quantities \emph{cannot} be viewed as spacetime metrics.

It is only in some special cases (\eg, ordinary nonlinear
electrodynamics) that the polynomial ${\cal P}_{4}(k)$ factorizes into
two quadratic forms,
\begin{equation}
\label{eq:factor}
{\cal P}_{4}(k)= 
\bigl(\gi_{(1)}^{\mu\nu} \; k_\mu \, k_\nu\bigr)  \; 
\bigl(\gi_{(2)}^{\alpha \beta} \; k_\alpha \,k_\beta \bigr),
\end{equation}
in which case we obtain two second-order dispersion relations:
\begin{equation}
\label{eq:disprel}
\gi_{(1)}^{\mu\nu} \; k_\mu k_\nu=0
\quad\mbox{and} \quad
\gi_{(2)}^{\mu\nu} \; k_\mu k_\nu=0,
\end{equation}
with momentum-independent matrices $\gi_{(r)}$.  We \emph{can} now
interpret these two matrices $\gi$ as the two (inverse) effective
metrics of a bi-metric theory. (More precisely they are representative
elements of two conformal classes of inverse metrics, since
multiplication by an arbitrary position-dependent scalar will not
modify the dispersion relations.)

In very special cases not only does (\ref{eq:factor}) hold, but also
$\gi_{(1)}^{\mu\nu}=\gi_{(2)}^{\mu\nu}$. (That is, the fourth-order
polynomial ${\cal P}_{4}(k)$ is a perfect square.) In this case one
ends up with a single quadratic dispersion relation of the familiar
form
\begin{equation}
\label{E:eomfc}
\gi^{\mu\nu}\;k_{\mu}\,k_{\nu}=0,
\end{equation}
where $\gi_{\mu\nu}$ is some symmetric tensor, which we shall call the
(inverse) effective metric (again defined only up to an arbitrary
conformal factor).

We now wish to investigate the conditions under which these
factorization (bi-metric) and uniqueness properties hold. Since the
entire formalism ultimately derives from the tensor
$\Omega^{\mu\alpha\nu\beta}$, we shall look for suitable algebraic
constraints on this tensor.

%------------------------------------------------------------------
\section{Single Effective metric}
%------------------------------------------------------------------
\label{S:effmet}
%------------------------------------------------------------------

In the case of a single effective metric we have
\begin{equation}
\label{E:eomfc2}
\gi^{\mu\nu}\,k_{\mu}\,k_{\nu}=0.
\end{equation}
It should be clear from our previous discussion that a necessary
condition for this to happen is the absence of birefringence.  We see
that the wave vector is now null with respect to this (unique)
``effective metric'' $\gi^{\mu\nu}$, which therefore defines an
effective geometry for the propagation of light.

We warn the reader that even when a unique (inverse) effective metric
$\gi^{\mu\nu}$ is defined, we always raise and lower indices using the
flat Minkowski metric $\eta_{\mu\nu}$, or in the presence of a
gravitational field the physical spacetime metric $g_{\mu\nu}$. The
effective metric itself, denoted $\g_{\mu\nu}$, is the matrix inverse
of $\gi^{\mu\nu}$. Because of the way indices are raised and lowered
using the physical metric you cannot use index placement to
distinguish $\gi = \g^{-1}$ from $\g$.

This single effective metric situation implies that (up to possibly a
piece proportional to the Levi--Civita tensor) the tensor
$\langle\psi|\Omega^{\mu\alpha\nu\beta}|\psi\rangle$ must be
algebraically constructible solely in terms of $\gi^{\mu\nu}$.  In
view of the symmetries of $\Omega^{\mu\alpha\nu\beta}$ we know,
without need for detailed calculation, that it must be of the form
\begin{equation}
\label{E:Omega2}
\langle\psi|\Omega^{\mu\alpha\nu\beta}|\psi\rangle =\Psi
\bigl(\gi^{\mu\nu} \;  \gi^{\alpha\beta} - 
      \gi^{\mu\beta} \;\gi^{\nu\alpha}\bigr)
+ \Phi \; \epsilon^{\mu\alpha\nu\beta}
\end{equation}
for some quantities $\Psi$ and $\Phi$. By appealing to the conformal
invariance of the null cones we can always absorb a factor of
$2\sqrt{|\Psi|}$ into the inverse metric $\gi$ and so rewrite this as
\begin{equation}
\label{E:Omega3}
\langle\psi|\Omega^{\mu\alpha\nu\beta}|\psi\rangle =
\pm{\bigl(\gi^{\mu\nu} \;  \gi^{\alpha\beta} - 
      \gi^{\mu\beta} \;\gi^{\nu\alpha}\bigr)
\over4}
+ \Phi \; \epsilon^{\mu\alpha\nu\beta}.
\end{equation}
Conversely, if $\Omega^{\mu\alpha\nu\beta}$ is of the form
(\ref{E:Omega3}), then the matrix (\ref{E:matrix}) is
\begin{eqnarray}
A^{\mu\nu}&\propto&
\Big\{ 
      \gi^{\mu\nu}
\bigl(\gi^{\alpha\beta}\,k_\alpha\,k_\beta\bigr) 
- 
\bigl(\gi^{\mu\alpha}\,k_\alpha\bigr) 
\;
\bigl(\gi^{\nu\beta}\,k_\beta\bigr)\Big\},
\nonumber \\
&&
\end{eqnarray}
and the photon propagation equation, $A^{\mu\nu}\,a_\nu=0$, becomes
\begin{equation}
\label{E:newpropageq} 
\bigl(\gi^{\alpha\beta}\,k_\alpha\,k_\beta\bigr)\gi^{\mu\nu}\,a_\nu-
\bigl(\gi^{\alpha\beta}\,a_\alpha\,k_\beta\bigr)\gi^{\mu\nu}\,k_\nu=0.
\end{equation}
This equation is obviously satisfied by the uninteresting gauge
modes $a\parallel k$, with no constraints on $k$. Solutions
corresponding to a non-vanishing $f^{\mu\nu}$ exist only if the
coefficient of $\gi^{\mu\nu}\,a_\nu$ is zero, \ie, if
(\ref{E:eomfc}) holds. Thus, the two polarization states
propagate with the same dispersion relation (\ref{E:eomfc}), and
there is no birefringence.

Substituting (\ref{E:eomfc}) back into the propagation equation
(\ref{E:newpropageq}) we find another relationship typical of this
case,
\begin{equation}
\label{E:gkepsilon} 
\gi^{\mu\nu}\,k_\mu\,a_\nu=0.
\end{equation}
Formally, the above equation looks like a gauge condition. This
might seem puzzling, because nowhere in the present subsection
have we fixed a gauge. In fact, (\ref{E:gkepsilon}) is a
consequence of the dynamical equation $A^{\mu\nu}\,a_\nu=0$, when
the ``on-shell'' condition (\ref{E:eomfc}) is satisfied, and it
does {\em not\/} imply any gauge fixing.

It is also interesting to notice that there is now a
self-consistency or ``bootstrap'' condition,
\begin{equation}
\label{E:bootstrap2}
\pm {3\over4}  \; \gi^{\mu\nu} =
\langle\psi|\Omega^{\mu\alpha\nu\beta}|\psi\rangle\; 
\g_{\alpha\beta}.
\end{equation}
We stress that these relations depend only on the assumed existence of
a single unique effective metric $\g_{\mu\nu}$ --- they do not make
any reference to other specifics of the external background fields
$B(x)$ or the quantum state.

Finally we point out that in this mono-refringent case the linearized
Lagrangian reduces to
\bee
{\cal L}_{\rm{linearized}} &=& 
\pm {1\over4} \; \gi^{\beta\mu} \;
f^\photon_{\mu\nu} \; \gi^{\nu\alpha} \;f^\photon_{\alpha\beta}
\nonumber\\
&&+
{1\over2} \Phi \; \epsilon^{\mu\nu\alpha\beta} 
\; f^\photon_{\mu\nu} \; f^\photon_{\alpha\beta}.
\eee
With hindsight, this is exactly what we should have expected. If we
now use this Lagrangian formulation to demand positivity of energy [a
feature missing from the purely kinematical analysis based on equation
(\ref{E:eom-master2})] then we should set $\pm \to +1$. Note that we
have also used the conformal invariance of the null cones to normalize
$\gi$ in the conventional manner. Finally the $\Phi$ term is simply
the well-known Pontryagin index.

%-------------------------------------------------------
\section{Ordinary Nonlinear Electrodynamics}
\label{app:schwinger}
%-------------------------------------------------------

In order to see how to develop a general ansatz that leads to
bi-metricity (perhaps not the most general ansatz) it is useful to
consider the explicit form of the tensor $\Omega_{\mu\nu\alpha\beta}$
for \emph{ordinary} nonlinear electrodynamics:
\begin{equation}
\label{E:Schwinger-lagrangian} 
\L_\NLE = \L\left({\cal F},{\cal G}\right).
\end{equation}
Here we have adopted the now common variables~\cite{Dittrich-Gies}
\begin{eqnarray}
\label{E:xy} 
{\cal F} &\equiv & \frac{1}{4} F_{\mu\nu} \;
F^{\mu\nu} = \frac{1}{2}\left(\vec{B}^2-\vec{E}^2\right),
\\
{\cal G} &\equiv & \frac{1}{4} F_{\mu\nu} \; \*F^{\mu\nu} =
-\vec{E}\cdot \vec{B}.
\end{eqnarray}
For such Lagrangians
\begin{eqnarray}
\label{E:Schwinger-like-omega} 
\Omega^{\mu\nu\alpha\beta} 
&=&
{1\over4}\;\left( \partial_{\cal F}\L \right) \; 
\big(
\eta^{\mu\alpha}\; \eta^{\nu_\beta} - \eta^{\mu\beta} \;
\eta^{\nu\alpha} 
\big)
\nonumber\\
&&+{1\over4}\;\left( \partial_{\cal G}\L \right)
\;\epsilon^{\mu\nu\alpha\beta}
\nonumber\\
&&
+F^{\mu\nu}\;F^{\alpha\beta}\; 
\left(\partial^{2}_{\cal F}\L\right)+
\*F^{\mu\nu}\;\*F^{\alpha\beta}\;\left(\partial^2_{\cal G}
\L\right)
\nonumber\\
&& +\left(F^{\mu\nu}\;\*F^{\alpha\beta}+
\*F^{\mu\nu}\;F^{\alpha\beta}\right)\partial_{\cal FG}\L.
\end{eqnarray}

As soon as one inserts this tensor into the photon equation of motion
(\ref{E:eom3}), the completely antisymmetric part proportional to the
Levi--Civita tensor drops out, because of the Bianchi identity
(\ref{E:bianchi2}). The remaining pieces reproduce the photon equation
of motion in the perhaps more usual form considered by Dittrich and
Gies~\cite{Dittrich-Gies}, or Novello and co-workers~\cite{NovelloPP}.

Unless one has a specific need to perform calculations to orders
higher than $O(\alpha^2)$, it is often sufficient to consider the
Euler--Heisenberg Lagrangian~\cite{Euler-Heisenberg} which, in the
${\cal F}$--${\cal G}$ formalism adopted above, takes the form
\begin{equation}
\label{E:ehxy} 
\L_{\rm EH}=
-\frac{1}{4\pi}\;{\cal F}+c_1\;{\cal F
}^2+c_2\;{\cal G}^2,
\end{equation}
with
\begin{equation}
\label{E:coeff} 
c_1=\frac{\alpha^2}{90\pi^2 m^4_\electron}, \qquad
c_2=\frac{7 \alpha^2}{360\pi^2 m^4_\electron}.
\end{equation}
The terms proportional to ${\cal F}^2$ and ${\cal G}^2$ of this
Lagrangian are quartic in the field, and describe the low-energy
limit of the box diagram in QED, when four photons couple to a
single virtual electron loop.  Thus, the Lagrangian
(\ref{E:ehxy}) is only accurate to order $\alpha^2$, and it is
meaningless to retain higher order terms within this model.
For the Euler--Heisenberg Lagrangian, the tensor
$\Omega^{\mu\nu\alpha\beta}$ is
\begin{eqnarray}
\label{E:MEH} \Omega^{\mu\nu\alpha\beta}&=&
\left(-{1\over16\pi}  + {c_1 \;{\cal F}\over 2} \right) \; 
\big(
\eta^{\mu\alpha}\; \eta^{\nu\beta} - \eta^{\mu\beta} \;\eta^{\nu\alpha} 
\big)
\nonumber\\
&& + {c_2 \; {\cal G}\over2} \; \epsilon^{\mu\nu\alpha\beta}
\nonumber\\
&&
+{c_1\over2} \; F^{\mu\nu}\;F^{\alpha\beta}
+{c_2\over2} \; \*F^{\mu\nu} \;\*F^{\alpha\beta}.
\end{eqnarray}
Again, when one inserts this tensor into the photon equation of motion
(\ref{E:eom3}), the completely antisymmetric part proportional to the
Levi--Civita tensor drops out.

Suppose we now adopt the ``rotated'' quantity
\be
K = \cos\theta \;F + \sin\theta \; \*F,
\ee
so that
\be
\*K = -\sin\theta \;F + \cos\theta \;  \*F.
\ee
Then for both (ordinary) NLE and Euler--Heisenberg electrodynamics we
can put $\Omega$ into the form
\bee
\label{E:ordinary} 
\Omega^{\mu\nu\alpha\beta}&=&
\Psi\; 
\big(
\eta^{\mu\alpha}\; \eta^{\nu\beta} - \eta^{\mu\beta} \;\eta^{\nu\alpha} 
\big)
\nonumber\\
&& + \Phi\; \epsilon^{\mu\nu\alpha\beta}
\nonumber\\
&&
+a \; K^{\mu\nu}\;K^{\alpha\beta}
+b\; \*K^{\mu\nu} \;\*K^{\alpha\beta}.
\eee
The matrix $A^{\mu\alpha}$ is then [suppressing explicit indices]
\bee
A &=& \Psi \{ \eta(k,k) \; \eta - (\eta k)\otimes(\eta k) \}
\nonumber\\
&& + a \; (K k)\otimes(K k) + b \; (\* K k)\otimes(\*K k).
\eee
Note that because both $K$ and $\*K$ are antisymmetric $K(k,k)$ and
$\*K(k,k)=0$; therefore $A \;k =0$ as expected. [We have defined
$\eta(k,k)\equiv \eta^{\mu\nu}\; k_\mu\; k_\nu$, $ K(k, k) \equiv
K^{\mu\nu} \; k_\mu \; k_\nu$, \etc ] Now consider $\det'(A)$, the
``reduced'' determinant in the three directions orthogonal to $k$. We
adopt this particular reduced determinant as a technically convenient
alternative to using the temporal gauge.  To be precise we define
\be
{\det}'(A) = \left.{\d[\det(A+\epsilon\,I)]\over \d\epsilon}\right|_{\epsilon=0}.
\ee
which implies
\be
{\det}'A={1 \over 3!}\big\{[{\rm Tr}(A)]^3+2[{\rm Tr}(A^3)]
-3[{\rm Tr}(A)][{\rm Tr}(A^2)]\big\}.
\ee
Now use this formula, or the fact that in 3 dimensions
\bee
&&\det{}_3(\lambda I + u\otimes u + v\otimes v) 
\\
&& \qquad 
= 
\lambda \; 
\left\{ 
\lambda^2 + \lambda (u^2 + v^2) + [u^2 v^2 - (u\cdot v)^2]
\right\}
\nonumber
\eee
(useful when $k$ is timelike with respect to $\eta$), to deduce
\bee
\det{}'(A) &=& - \Psi \; I(k,k) \;
\Big\{ 
\Psi^2 \; \eta(k,k)^2 
\nonumber\\
&&
- \Psi\; \eta(k,k)\; [a \;\gamma(K k, K k) + b \; \gamma(\*K k, \*K k) ] 
\nonumber\\
&&
+ a b [\gamma(K k, K k)\; \gamma(\*K k, \*K k) 
\nonumber\\
&&
\qquad
- \gamma(K k,\*K k)^2]
\Big\}.
\eee
[The presence of the $-$ sign arises from the indefinite nature of the
metric $\eta$. We have also verified the above formulae via explicit
evaluation of the determinants using Maple. Note that we now
distinguish between the (inverse) contravariant Minkowski metric
$\eta$ and the covariant Minkowski metric $\gamma =\eta^{-1}$. The
wavevector $k$ is always taken to be covariant while both $K$ and
$\*K$ are assumed doubly contravariant. We have defined $\gamma(K k, K
k) \equiv \gamma_{\mu\nu} \; (K k)^\mu \; (K k)^\nu$, \etc ]

Discarding the uninteresting factor of $I(k,k) = k^T k$ (it
corresponds to the factor $\omega^2$ encountered when we worked in
temporal gauge) we identify
\bee
{\cal P}_4(k) &=& 
\Psi^2\; \eta(k,k)^2 
\nonumber\\
&&
- \Psi\;\eta(k,k)\; [a \;\gamma(K k, K k) + b \; \gamma(\*K k, \*K k) ] 
\nonumber\\
&&
+ a b [\gamma(K k, K k)\; \gamma(\*K k, \*K k) 
\nonumber\\
&&
\qquad
- \gamma(K k,\*K k)^2].
\eee
This is clearly a quartic in $k$, and the ``miracle'' of (ordinary)
nonlinear electrodynamics is that it factorizes into two
quadratics. To establish this factorization we use some very special
properties of (3+1) dimensions:
\bee
T_\Maxwell 
&=& F \; \gamma \; F - {1\over4}  \tr(F \gamma F \gamma) \; \eta
\\
&=& \*F \; \gamma \; \*F - {1\over4} \tr(\*F \gamma \*F \gamma) \; \eta
\\
&=& \*F \; \gamma \; \*F + {1\over4} \tr(F \gamma F \gamma) \; \eta
\eee
and
\bee
F \;\gamma \;\* F = {\cal G} \; \eta = \*F \;\gamma \;F.
\eee
In terms of the ``rotated variables'' the Maxwell stress energy tensor
is given by
\bee
T_\Maxwell 
&=& K \;\gamma \;K - {1\over4} \tr(K \gamma K \gamma) \;\eta
\\
&=& \*K \;\gamma \;\*K - {1\over4} \tr(\*K \gamma \*K \gamma) \;\eta
\\
&=& \*K \;\gamma \;\*K + {1\over4} \tr(K \gamma K \gamma) \;\eta.
\eee
Additionally
\bee
K \;\gamma \;\* K = {\cal G}_\theta \;\eta = \*K \;\gamma \;K,
\eee
where
\bee
{\cal F}_\theta &=&    \cos(2\theta) \;{\cal F} +\sin(2\theta) \;{\cal G}
\\
{\cal G}_\theta &=&  - \sin(2\theta) \;{\cal F} +\cos(2\theta) \;{\cal G}.
\eee
{From} these relations we can deduce
\bee
\gamma(K k, K k) &=& - T_\Maxwell(k,k) - {\cal F}_\theta \; \eta(k,k),
\\
\gamma(\*K k, \*K k) &=& - T_\Maxwell(k,k) + {\cal F}_\theta \; \eta(k,k),
\\
\gamma(K k, \*K k) &=& - {\cal G}_\theta \; \eta(k,k).
\eee
When substituted into ${\cal P}_4$ this implies (dropping the explicit
``$\Maxwell$'' subscript)
\be
{\cal P}_4 = 
a_0 \; \eta(k,k)^2 + a_1 \; \eta(k,k) \; T(k,k) + a_2 \; T(k,k)^2,
\ee
for suitable $a_0$, $a_1$, $a_2$. Indeed
\bee
a_0 &=& 
\Psi^2 + \Psi (a-b) {\cal F}_\theta 
- a b \left( {\cal F}_\theta{}^2 +  {\cal G}_\theta{}^2 \right),
\\
a_1 &=&
\Psi (a+b),
\\
a_2 &=&
a b.
\eee
The quartic will factorize provided
\be 
a_0 + a_1 x + a_2 x^2 =0
\ee
has a solution in the real numbers. Fortunately the discriminant is
easily seen to be a sum of squares and so is always positive.
\be
a_1^2-4\,a_0\,a_2 = 
\left[\Psi(a-b)+2\,ab{\cal F}_\theta\right]^2 + \left[2\,ab\,{\cal G}_\theta\right]^2 
\geq 0.
\ee
This is now enough to guarantee that ${\cal P}_4$ factorizes, indeed
\be
{\cal P}_4 = a_0 
\left[\eta(k,k)+ b_1 \; T(k,k)\right]\; 
\left[\eta(k,k)+ b_2 \; T(k,k)\right] 
\ee
for suitable $a_0$, $b_1$, $b_2$. These coefficients will be functions
of $\Psi$, $a$, $b$, ${\cal F}_\theta$ and ${\cal G}_\theta$ whose
precise form is not needed for the point we are currently making:
{\emph{As long as the Lagrangian is only a function of the two
invariants ${\cal F}$ and ${\cal G}$, then the theory is not just
birefringent, it is truly bi-metric.}}

%------------------------------------------------------
\section{A general bi-metric ansatz}
%------------------------------------------------------
\label{S:general bimetric ansatz}
%------------------------------------------------------

Based on the above we can now guess a general ansatz for
$\Omega^{\mu\nu\alpha\beta}$ that always leads to bi-metric
propagation (we do not guarantee that this is the most general
ansatz). Let
\bee
\label{E:ansatz} 
\Omega^{\mu\nu\alpha\beta}&=&
\Psi\; 
\big(
g^{\mu\alpha}\; g^{\nu\beta} - g^{\mu\beta} \;g^{\nu\alpha} 
\big)
\nonumber\\
&& + \Phi\; \epsilon^{\mu\nu\alpha\beta}
\nonumber\\
&&
+a \; K^{\mu\nu}\;K^{\alpha\beta}
+b\; \*K^{\mu\nu} \;\*K^{\alpha\beta},
\eee
where $g^{\mu\nu}$ is now any symmetric matrix of Lorentzian signature
and $K^{\mu\nu}$ is any anti-symmetric matrix --- in particular $K$
does not necessarily have anything to do with the electromagnetic
field. All the analysis in the previous section can now be converted
into purely algebraic statements about $g$ and $K$. For example $T$ no
longer has the interpretation of being the Maxwell stress-energy
tensor, it is simply an algebraic matrix defined by
\be
T = K g^{-1} K - {1\over4}\tr(K g^{-1} K g^{-1}) g.
\ee
Reinterpreting everything in this purely algebraic manner, and
repeating the analysis of the previous section, we see that
(\ref{E:ansatz}) leads to bi-metric propagation with the two light
cones being given by linear combinations of $g$ and $T$.

Though rather general, this is not likely to be the most general
bi-metric ansatz. To see this note that (\ref{E:ansatz}) above appears
to contain $10(g)+6(K)+4(\Psi,\Phi,a,b)=20$ free parameters. But
$\Psi$ can be absorbed by redefining $g$, while $\Phi$ does not affect
the equations of motion, $a$ can be absorbed by redefining $K$, and an
overall scale factor does not affect the equations of motion. This
leaves 16 physically interesting free parameters in (\ref{E:ansatz})
to be compared with $2\times9=18$ free parameters encoded in a pair of
light cones. We have not as yet been able to deduce the most general
form of $\Omega^{\mu\nu\alpha\beta}$ compatible with bi-metricity.

In terms of the linearized Lagrangian our bi-metric ansatz
(\ref{E:ansatz}) corresponds to
\bee
{\cal L}_{\rm{linearized}} &=& 
{1\over2} \; \Psi \tr(f_\photon \;g^{-1} \;f_\photon \;g^{-1}) 
\nonumber\\
&+&
{1\over2} \Phi \tr(f_\photon \;g^{-1} \;\*f_\photon \;g^{-1}) 
\nonumber\\
&+&
{1\over2} \;  a \tr(K \;g^{-1}\;f_\photon \;g^{-1})^2 
\nonumber\\
&+& {1\over2} \; b  \tr(\*K  \;g^{-1}\; f_\photon \;g^{-1})^2.
\eee
%

%------------------------------------------------------
\section{Birefringence without bi-metricity}
%------------------------------------------------------
\label{S:birefringence-without-bimetricity}
%------------------------------------------------------

To wrap up, can we now give a simple explicit example of a model that
is birefringent \emph{without} being bi-metric? Suppose we have a pair
of two-forms $J_1$ and $J_2$ with $J_1\not\propto \*J_2$. Then
consider (as a particular example)
\bee
\label{E:biref} 
\Omega^{\mu\nu\alpha\beta}&=&
\Psi\; 
\big(
\eta^{\mu\alpha}\; \eta^{\nu\beta} - \eta^{\mu\beta} \;\eta^{\nu\alpha} 
\big)
\nonumber\\
&& + \Phi\; \epsilon^{\mu\nu\alpha\beta}
\nonumber\\
&&
+a \; J_1^{\mu\nu}\;J_1^{\alpha\beta}
+b\;  J_2^{\mu\nu} \; J_2^{\alpha\beta}.
\eee
In this case, because you are now not satisfying the special algebraic
constraints of the previous section, there is no reason for the
determinant to factorize. Indeed
\bee
{\cal P}_4(k) &=& 
\Psi^2\; \eta(k,k)^2 
\nonumber\\
&&
- \Psi\;\eta(k,k)\; [a \;\gamma(J_1 k, J_1 k) + b \; \gamma(J_2 k, J_2 k) ] 
\nonumber\\
&&
+ a b [\gamma(J_1 k, J_1 k)\; \gamma(J_2 k, J_2 k) 
\nonumber\\
&&
\qquad
- \gamma(J_1 k,J_2 k)^2].
\eee
In terms of the linearized Lagrangian (suppressing the inverse
Minkowski metric $\gamma=\eta^{-1}$ as being understood)
\bee
{\cal L}_{\rm{linearized}} &=& 
{1\over2} \; \Psi \tr(f_\photon^2) + {1\over2} \Phi \tr(f_\photon \*f_\photon) 
\nonumber\\
&+&
{1\over2} \; \left\{ a \tr(J_1 \;f_\photon)^2 + b \tr(J_2 \; f_\photon)^2 \right\}.
\nonumber\\
&&
\eee
The key point here is that the linearized Lagrangian is \emph{not} a
simply a function of the invariants ${\cal F}$ and ${\cal G}$. In
going to generalized nonlinear electrodynamics we have permitted
additional structure in the form of external fields and boundary
conditions. Unless these external fields satisfy rather particular
conditions (\eg, in the present example, $J_1\propto \* J_2$) there is
no reason to believe the Fresnel determinant factorizes, and no reason
to expect a bi-metric theory.

Indeed, if one picks a tensor $\Omega$ ``at random'' and explicitly
evaluates $P_4(k)$ (using a symbolic program such as Maple) one
rapidly concludes that arranging factorization (and so bi-metricity)
is not an easy task. Bi-metricity is not generic in the set of all
birefringent theories.

%------------------------------------------------------
\section{Conclusions}
%------------------------------------------------------
\label{S:conclusions}
%------------------------------------------------------

In this Festschrift article we have discussed, on quite general
grounds, the phenomenon of bi-refringence and bi-metricity in
[generalized] (3+1)-dimensional non-linear electrodynamics.  Our
treatment encompass any non-linear electrodynamic theory in
interaction with an arbitrary number of external non-dynamical fields
characterizing a general medium (for example, a flowing dielectric,
external gravitational fields, moving Casimir plates, \etc). We have
seen that the phenomenon of bi-refringence is both generic and easily
established in these theories.  In (3+1) dimensions, the Fresnel
equation is quartic and in special cases (\eg, ordinary nonlinear
electrodynamics) factorizes into two quadratics (\ie, two metrics),
typically different from each other.  In more specialized situations
these two metrics can be identical, leaving no opportunity for
bi-refringence.

If we consider generalized nonlinear electrodynamics, then
because of the presence of additional background fields, the close
link between bi-refringence and bi-metricity can be broken ---
in such situations the Fresnel equation is intrinsically
\emph{quartic} and to naturally describe the geometry one would need
to go beyond the notion of Lorentzian geometry and instead
introduce the notion of a pseudo--Finsler geometry as described
in~\cite{normal-modes2}.

In closing we emphasise that the use of nonlinear extensions to
electrodynamics is currently becoming ubiquitous. The implied notions
of birefringence, bi-metricity, and pseudo-Finsler geometries will
doubtless continue to attract considerable attention. Professor
Novello's work on effective geometries and birefringence will continue
to have important repercussions down the road.

%-------------------------------------------------
\section*{Acknowledgements}
%-------------------------------------------------

The authors would like to thank C. Herdeiro for bringing
reference~\cite{Gibbons2} to our attention. Additionally, MV and CB
would like to thank Professor {\Mario} Novello for his hospitality
during the workshop on ``Analog models of General Relativity''.

%========================================================================
% The references have proper Spires citations attached.
% This is supposed to help the Spires staff in updating their database.
% Don't muck around with this unless you know what you are doing.
% Don't touch the commented ``citation = '' lines.
%========================================================================

%----------------------------------------------------------------------------

\begin{thebibliography}{99}
%----------------------------------------------------------------


%-------------------------------------------------------------------------
\bibitem{Born-Infeld}
M.~Born and L.~Infeld,
``Foundations Of The New Field Theory,''
Proc.~Roy.~Soc.~Lond.~A {\bf 144} (1934) 425.
%%CITATION = PRSLA,A144,425;%%

%----------------------------------------------------------------
\bibitem{Euler-Heisenberg}
W.~Heisenberg and H.~Euler, ``Consequences of Dirac's theory of
positrons,'' Z.~Phys. {\bf 98}, 714--732 (1936).
%%CITATION = ZEPYA,98,714;%%

%----------------------------------------------------------------
\bibitem{Schwinger}
J.~Schwinger, ``On gauge invariance and vacuum polarization,''
Phys.~Rev.~ {\bf 82}, 664--679 (1951).
%%CITATION = PHRVA,82,664;%%

%----------------------------------------------------------------
\bibitem{EMbackground}
S.~L.~Adler, 
``Photon splitting and photon dispersion in a strong
magnetic field,'' Ann.~Phys.~(N.Y.) {\bf 67}, 599--647 (1971);
%%CITATION = APNYA,67,599;%%
\\
%S.~L.~Adler,
``Comment on ``Photon splitting in strongly magnetized objects
revisited'','' astro-ph/9601156.
%%CITATION = ASTRO-PH 9601156;%%
\\
S.~L.~Adler and C.~Schubert, ``Photon splitting in a strong
magnetic field: Recalculation and comparison with previous
calculations,'' Phys.~Rev.~Lett.~ {\bf 77}, 1695--1698 (1996)
[hep-th/9605035].
%%CITATION = HEP-TH 9605035;%%
\\
E.~Brezin and C.~Itzykson,
``Polarization Phenomena In Vacuum Nonlinear Electrodynamics,''
Phys.~Rev.~D {\bf 3} (1971) 618.
%%CITATION = PHRVA,D3,618;%%
%----------------------------------------------------------------

\bibitem{Gbackground}
I.~T.~Drummond and S.~J.~Hathrell, 
``QED vacuum polarization in a
background gravitational field and its effect on the velocity of
photons,'' Phys.~Rev.~D {\bf 22}, 343--355 (1980).
%%CITATION = PHRVA,D22,343;%%
\\
%----------------------------------------------------------------
V.~M.~Mostepanenko and N.~N.~Trunov, {\em The Casimir Effect and
its Applications\/} (Oxford Science Publications, Clarendon
Press, Oxford, 1997).
%%CITATION = NONE;%%

%-------------------------------------------------------------------------
\bibitem{GBCbackground}
H.~B.~Casimir, ``On the attraction between two perfectly
conducting plates,'' Kon.~Ned.~Akad.~Wetensch.~Proc.~ {\bf 51},
793--795 (1948).
%%CITATION = KNAWA,51,793;%%
\\
%-------------------------------------------------------------------------
%\bibitem{Casimir-report}
G.~Plunien, B.~Muller and W.~Greiner, ``The Casimir effect,''
Phys.~Rep.~{\bf 134}, 87--193 (1986).
%%CITATION = PRPLC,134,87;%%
\\
%----------------------------------------------------------------
%\bibitem{Sch90}
K.~Scharnhorst, ``On propagation of light in the vacuum between
plates,'' Phys.~Lett.~{\bf B236}, 354--359 (1990).
%%CITATION = PHLTA,B236,354;%%
\\
%----------------------------------------------------------------
%\bibitem{Bar90}
G.~Barton, ``Faster-than-$c$ light between parallel mirrors. The
Scharnhorst effect rederived,'' Phys.~Lett.~{\bf B237}, 559--562
(1990).
%%CITATION = PHLTA,B237,559;%%
\\
%-------------------------------------------------------------------------
%\bibitem{SchBar93}
G.~Barton and K.~Scharnhorst, ``QED between parallel mirrors:
light signals faster than $c$, or amplified by the vacuum,'' J.\
Phys.~A {\bf 26}, 2037--2046 (1993).
%%CITATION = JPAGB,A26,2037;%%
\\
%----------------------------------------------------------------
%\bibitem{Sch98}
K.~Scharnhorst, ``The velocities of light in modified QED vacua,''
Ann.~Phys.~(Leipzig) {\bf 7}, 700--709 (1998) [hep-th/9810221].
%%CITATION = HEP-TH 9810221;%%

%----------------------------------------------------------------
\bibitem{Addbackground}
%\bibitem{DS94}
R.~D.~Daniels and G.~M.~Shore, `` `Faster than light' photons and
charged black holes,'' Nucl. Phys. {\bf B425}, 634--650 (1994)
[hep-th/9310114].
%%CITATION = HEP-TH 9310114;%%
\\
%--------------------------------------------------------------------
%\bibitem{DS96}
R.~D.~Daniels and G.~M.~Shore, `` `Faster than light' photons and
rotating black holes,'' Phys. Lett.  {\bf B367}, 75--83 (1996)
[gr-qc/9508048].
%%CITATION = GR-QC 9508048;%%
\\
%--------------------------------------------------------------------
%\bibitem{Shore96}
G.~M.~Shore, `` `Faster than light' photons in gravitational
fields --- Causality, anomalies and horizons,'' Nucl.~Phys.~{\bf
B460}, 379--394 (1996) [gr-qc/9504041].
%%CITATION = GR-QC 9504041;%%
\\
%--------------------------------------------------------------------
%\bibitem{Latorre}
J.~I.~Latorre, P.~Pascual and R.~Tarrach, ``Speed of light in
non-trivial vacua,'' Nucl.~Phys.~{\bf B437}, 60--82 (1995)
[hep-th/9408016].
%%CITATION = HEP-TH 9408016;%%
\\
%----------------------------------------------------------------
%\bibitem{apparent-superluminal}
A.~D.~Jackson, A.~Lande and B.~Lautrup, ``Apparent superluminal
behavior,'' physics/0009055.
%%CITATION = PHYSICS 0009055;%%
\\
%----------------------------------------------------------------
%\bibitem{no-thing}
A.~M.~Steinberg, ``No thing goes faster than light,'' Phys.~World
{\bf 13} (9), 21 (2000).
%%CITATION = NONE;%%

%----------------------------------------------------------------
\bibitem{NovelloGS}
M.~Novello, S.~E.~Perez Bergliaffa and J.~M.~Salim,
``Singularities in General Relativity coupled to nonlinear electrodynamics,''
Class.~Quant.~Grav.~ {\bf 17} (2000) 3821
[arXiv:gr-qc/0003052].
%%CITATION = GR-QC 0003052;%%
\\
%--------------------------------------------------------------
M.~Novello, J.~M.~Salim, V.~A.~De Lorenci and R.~Klippert,
``Effective Lagrangian for electrodynamics and avoidance of the singular origin of the universe,''
arXiv:gr-qc/9806076.
%%CITATION = GR-QC 9806076;%%

%----------------------------------------------------------------
\bibitem{NovelloPP}
V.~A.~De Lorenci, R.~Klippert, M.~Novello and J.~M.~Salim,
``Light propagation in non-linear electrodynamics,''
Phys.~Lett.~B {\bf 482} (2000) 134
[arXiv:gr-qc/0005049].
%%CITATION = GR-QC 0005049;%%
\\
%----------------------------------------------------------------
M.~Novello, V.~A.~De Lorenci, J.~M.~Salim and R.~Klippert,
``Geometrical aspects of light propagation in nonlinear electrodynamics,''
Phys.~Rev.~D {\bf 61} (2000) 045001
[arXiv:gr-qc/9911085].
%%CITATION = GR-QC 9911085;%%
\\
%----------------------------------------------------------------
M.~Novello and J.~M.~Salim,
``Effective Electromagnetic Geometry,''
Phys.~Rev.~D {\bf 63} (2001) 083511.
%%CITATION = PHRVA,D63,083511;%%

%----------------------------------------------------------------
\bibitem{Plebansky}
J. Plebansky, in {\it Lectures on non-linear electrodynamics},
(Ed. Nordita, Copenhagen, 1968).

%----------------------------------------------------------------
\bibitem{NovelloBH}
%\cite{Novello:1998rr}
%\bibitem{Novello:1998rr}
M.~Novello, V.~A.~De Lorenci and E.~Elbaz,
``Can non gravitational black holes exist?,''
CBPF-NF-032-98.
\\
%----------------------------------------------------------------
M.~Novello, V.~A.~De Lorenci and E.~Elbaz,
``Can purely electromagnetic interaction produce black holes?,''
arXiv:gr-qc/9702054.
%%CITATION = GR-QC 9702054;%%

%----------------------------------------------------------------
\bibitem{NovelloWH}
F.~Baldovin, M.~Novello, S.~E.~Perez Bergliaffa and J.~M.~Salim,
``A non-gravitational wormhole,''
Class.~Quant.~Grav.~ {\bf 17} (2000) 3265
[arXiv:gr-qc/0003075].
%%CITATION = GR-QC 0003075;%%

%----------------------------------------------------------------
\bibitem{NovelloCTC}
%\cite{Novello:2001fv}
%\bibitem{Novello:2001fv}
M.~Novello, J.~M.~Salim, V.~A.~De Lorenci and E.~Elbaz,
``Nonlinear electrodynamics can generate a closed spacelike path for photons,''
Phys.~Rev.~D {\bf 63} (2001) 103516.
%%CITATION = PHRVA,D63,103516;%%
\\
%----------------------------------------------------------------
M.~Novello, V.~A.~De Lorenci, E.~Elbaz and J.~M.~Salim,
``Closed lightlike curves in non-linear electrodynamics,''
arXiv:gr-qc/0003073.
%%CITATION = GR-QC 0003073;%%

%-------------------------------------------------------------------------
\bibitem{Dittrich-Gies}
W.~Dittrich and H.~Gies, ``Light propagation in nontrivial QED
vacua,'' Phys.~Rev.~D {\bf 58}, 025004 (1998) [hep-ph/9804375].
%%CITATION = HEP-PH 9804375;%%
\\
%----------------------------------------------------------------
W.~Dittrich and H.~Gies, {\em Probing the Quantum Vacuum\/}, Springer
Tracts in Modern Physics {\bf 166} (Springer, 2000)
%%CITATION = NONE;%%

%-------------------------------------------------------------------------
\bibitem{Schrodinger}
E. \Schrodinger, 
{\em Contributions to Born's New Theory of the Electromagnetic Field}, 
Proc. Roy. Soc.  {\bf 150A} (1935) 465. 
\\
E. \Schrodinger, {\em Non-Linear Optics}, 
Proc. Roy. Irish. Acad. {\bf A 47} (1942) 77.
\\
E. \Schrodinger, 
{\em A new exact solution in non-linear optics (two-wave-system)}, 
Proc. Roy. Irish. Acad. {\bf A 49} (1943) 59.

%----------------------------------------------------------------
\bibitem{Boillat}
G. Boillat, 
{\em Vari\'et\'es caract\'eristiques ou surfaces d'ondes en \'electrodynamiques non lin\'eaire}, 
C. R. Acad. Sci. Paris {\bf 262} (1966) 1285. 
\\
G. Boillat, 
{\em Surfaces d'ondes compar\'es de la theorie
d'Einstein-Schroedinger et de l'\'electrodynamique
non lin\'eaire; champs absolus}, 
C. R. Acad. Sci. Paris {\bf 264} (1967) 113.
\\
G. Boillat, 
{\em Nonlinear Electrodynamics: Lagrangians and Equations of Motion}, 
J. Math. Phys. {\bf 11} (1970) 941.    
\\
G. Boillat, {\em Exact Plane-Wave Solution of Born-Infeld Electrodynamics}, 
Lett. al Nuovo Cimento {\bf 4} (1972) 274.
\\
G. Boillat, 
{\em Shock relations in Non-Linear Electrodynamics}, 
Phys. Lett. {\bf 40A} (1972) 9.
\\
G. Boillat, 
{\em Convexit\'e et hyperbolicit\'e en \'electrodynamique non-lin\'eaire}, 
C. R. Acad. Sci. Paris {\bf 290} (1980) 259.
%%CITATION = NONE;%%

%--------------------------------------------------------------
\bibitem{normal-modes2}
C. \Barcelo, S. Liberati, and M. Visser,
``Refringence, field theory, and normal modes'',
gr-qc/0111059.
%%CITATION = GR-QC 0111059;%%

%----------------------------------------------------------------
\bibitem{Barcelo}
C.~\Barcelo, S.~Liberati, and M.~Visser,
``Analog gravity from Bose--Einstein condensates'',
Class.~Quant.~Grav.~ {\bf 18} (2001) 1137
[gr-qc/0011026];
%%CITATION = GR-QC 0011026;%%

%--------------------------------------------------------------
\bibitem{normal-modes}
C. \Barcelo, S. Liberati, and M. Visser,
``Analog gravity from field theory normal modes?'',
Classical and Quantum Gravity {\bf 18} (2001) 3595-3610 [gr-qc/0104001]; 
%%CITATION = GR-QC 0104001;%%
\\
%--------------------------------------------------------------
%\bibitem{normal-modes-emergent}
C. \Barcelo, S. Liberati, and M. Visser,
``Einstein gravity as an emergent phenomenon?'',
gr-qc/0106002; International Journal of Modern Physics (in press).
%%CITATION = GR-QC 0106002;%%

%--------------------------------------------------------------
\bibitem{Polchinski}
J.~Polchinski,
``String Theory. Vol. 1: An Introduction To The Bosonic String,''
{\it  Cambridge, UK: Univ. Pr. (1998) 402 p}.

%--------------------------------------------------------------
\bibitem{Gibbons1}
G.~W.~Gibbons,
``Aspects of Born-Infeld Theory and String/M-Theory,''
arXiv:hep-th/0106059.
%%CITATION = HEP-TH 0106059;%%
\\
%--------------------------------------------------------------
G.~W.~Gibbons,
``Born-Infeld particles and Dirichlet p-branes,''
Nucl.~Phys.~B {\bf 514} (1998) 603
[arXiv:hep-th/9709027].
%%CITATION = HEP-TH 9709027;%%

%--------------------------------------------------------------
\bibitem{Gibbons2}
G.~W.~Gibbons and C.~A.~Herdeiro,
``Born-Infeld theory and stringy causality,''
Phys.~Rev.~D {\bf 63} (2001) 064006
[arXiv:hep-th/0008052].
%%CITATION = HEP-TH 0008052;%%

%---------------------------------------------------------------
\bibitem{Oblique}
S.~Liberati, S.~Sonego and M.~Visser,
``Scharnhorst effect at oblique incidence,''
Phys.\ Rev.\ D {\bf 63} (2001) 085003
[arXiv:quant-ph/0010055].
%%CITATION = QUANT-PH 0010055;%%

%----------------------------------------------------------------
\bibitem{Landau}
L.~D.~Landau, E.~M.~Lifshitz and L.~P.~Pitaevskii, {\em
Electrodynamics of Continuous Media\/}, 2nd edition (Oxford,
Pergamon Press, 1984).
%%CITATION = NONE;%%

%----------------------------------------------------------------
\bibitem{Hadamard}
J.~Hadamard, 
{\em Le\c{c}ons sur la propagation des ondes et les 
\'equations de l'hydrodynamique}, 
(Hermann, Paris, 1903).
%%CITATION = NONE;%%

%----------------------------------------------------------------

%----------------------------------------------------------------
\end{thebibliography}
\end{document}